\documentclass[preprint,12pt]{elsarticle}

\usepackage{latexsym}
\usepackage{epsfig}
\usepackage{graphicx}
\usepackage{amsmath}
\usepackage{hyperref}

\usepackage[lmargin=2cm,rmargin=2cm,tmargin=2.5cm,bmargin=3.5cm]{geometry}

\journal{Physics Letters B}

\begin{document}

\begin{frontmatter}

\title{Nuclear shadowing in photoproduction of $\rho$ 
mesons in ultraperipheral nucleus collisions at RHIC and the LHC} 

\author[label1]{L. Frankfurt}

\author[label2]{V. Guzey}

\author[label3]{M. Strikman}

\author[label2]{M. Zhalov}

\address[label1]{Nuclear Physics Department, School of Physics and Astronomy, Tel Aviv
University, 69978 Tel Aviv, Israel}

\address[label2]{National Research Center ``Kurchatov Institute'', Petersburg Nuclear Physics Institute (PNPI), Gatchina, 188300, Russia}

\address[label3]{Department of Physics, the Pennsylvania State University, State College, PA 16802, USA}


\begin{abstract}

We argue that with an increase of the collision energy, elastic photoproduction of $\rho$ mesons on nuclei becomes
affected by the significant cross section of photon inelastic diffraction
into large masses, 
which results in the sizable inelastic nuclear shadowing correction to $\sigma_{\gamma A \to \rho A}$ and 
the reduced effective $\rho$--nucleon cross section.  We take these effects into account by  combining the 
vector meson dominance model,
which we upgrade to include the contribution of high-mass fluctuations
of the photon according to QCD constraints,
and the Gribov--Glauber approximation for nuclear shadowing, where the inelastic nuclear shadowing is 
included by means of cross section fluctuations. The resulting approach allows us to successfully describe the 
data on elastic $\rho$ photoproduction on nuclei in heavy ion UPCs in the 
$7 \ {\rm GeV} < W_{\gamma p} < 46$ GeV 
energy range and to predict the value of the
cross section of coherent $\rho$ photoproduction in Pb-Pb UPCs 
at $\sqrt{s_{NN}}=5.02$ TeV in Run 2 at the LHC, 
$d\sigma_{Pb Pb \to \rho Pb Pb} (y=0)/dy= 560 \pm 25$ mb.

\end{abstract}

\begin{keyword}

Ion-ion ultraperipheral collisions \sep exclusive photoproduction of $\rho$ mesons 
\sep nuclear shadowing \sep vector meson dominance model

\end{keyword}

\end{frontmatter}

\section{Introduction}
\label{sec:Intro}

At high photon beam energies $E_{\gamma}$,  the photon 
participates in the strong interaction with hadrons through its fluctuation into hadronic components.
The lifetime of 
these fluctuations is characterized by the coherence length $l_c =2E_{\gamma}/M^2$, where $M$ is the mass of a given 
hadronic component.
With an increase of $E_{\gamma}$, $l_c$ increases and becomes larger 
than the target size $R_T$ for progressively heavier hadronic fluctuations of the photon, 
which is manifested in an increase of photon inelastic diffraction into large masses.
This means that the photon can be represented as a coherent superposition of hadronic
fluctuations interacting with the target with a wide spectrum of cross sections. 
This picture can be implemented in terms of either hadronic 
(in particular, vector mesons) or quark--gluon degrees of freedom.

In the 60's and early 70's, it was suggested that the observed hadron-like 
behavior of a photon in photon--hadron 
interactions can be represented by the vector meson dominance model (VMD), 
which assumes that the photon fluctuates 
into $\rho$, $\omega$ and $\phi$ mesons that subsequently interact with 
hadrons~\cite{Feynman:1973xc,Bauer:1977iq}.
The VMD model 
successfully explained the behavior of the pion form factor, certain features of the nucleon form factors at 
small momentum transfers and the major part ($\approx 80$\%) of the 
real photon--nucleon total cross section $\sigma_{\gamma p}$. Combining VMD, 
the constituent quark model~\cite{Levin:1965mi} 
and the Regge--Gribov theory of high energy hadron--hadron scattering,
the 
soft Pomeron phenomenology had been developed and actively used~\cite{Donnachie:1992ny,Donnachie:1994zb}  
to study properties of the vector meson interaction with nucleons in light vector meson 
photoproduction and electroproduction at 
small photon virtualities $Q^2$.
One of the key features of this approach 
is the assumption that $\sigma_{\rho N}=\sigma_{\pi N}$, which is based on the additive quark model 
 ($\sigma_{\rho N}$ and $\sigma_{\pi N}$ are the total $\rho$--nucleon and pion--nucleon cross sections, 
 respectively). In a wide range of pion energies, the total pion--nucleon cross section is described well by a 
sum of the soft Pomeron and the secondary Reggeon exchanges~\cite{Donnachie:1992ny,Donnachie:1994zb} 
(we refer to this model as DL94).

An increase of the photon virtuality $Q^2$ leads to a gradual 
transition from the soft nonperturbative QCD regime to 
the perturbative one, which is clearly revealed in HERA measurements 
of vector meson electroproduction on the proton at high energies,
for the review and references, see, e.g.~\cite{Ivanov:2004ax}. 
To explain in a consistent way the $\approx 20\%$ discrepancy between the experimental
value of $\sigma_{\gamma p}$ and VMD predictions 
and the behavior of photon-induced processes with an increase of $Q^2$, new theoretical 
approaches have been developed. 
On the one hand, within the framework of hadronic description, the VMD model was generalized 
on the basis of the mass dispersion relation 
to include higher-mass resonances with
their diagonal and non-diagonal transitions~\cite{Gribov:1968gs, Sakurai:1972wk, Fraas:1974gh,Ditsas:1976yv};
the resulting approach is referred to as the generalized vector meson dominance (GVMD) model.
On the other hand, in the 
QCD framework,
the photon wave function in the strong interaction can be modeled
as superposition of quark--antiquark pairs (dipoles)
and higher Fock states, which interact with the target. 
The approach in which one treats the photon as a quark--antiquark pair 
  is called the color dipole model (CDM).
  
From the point of view of the quark--hadron duality,
these two approaches should be in principle equivalent, but this
equivalence is destroyed in their practical realization. 
In particular, to apply the CDM in the nonperturbative domain,
for example, for the description of photoproduction of light vector mesons, 
the approach should be generalized to take into account more complicated than $q\bar q$ states of large masses
and also to model the cross section of the interaction of large-size dipoles with
nucleons. On the other hand, the GVMD model includes coupling constants of the photon to
higher-mass resonances and amplitudes of their diagonal and non-diagonal
transitions, which are in general unknown. As a result, both approaches
require to use phenomenology and engage additional experimental information.

Coherent photoproduction of light vector mesons on nuclear targets at low and intermediate energies has been 
considered as theoretically well understood within the framework of the VMD model and the 
Glauber theory of multiple scattering~\cite{Glauber:1977nj} 
taking into account the finite coherence length
(for brevity, we refer to the resulting approach as VMD-GM).
Recently the ALICE collaboration presented results 
on exclusive $\rho$ meson production at the central rapidity in Pb-Pb ultraperipheral
collisions (UPCs) at $\sqrt{s_{NN}}=2.76$ TeV~\cite{Adam:2015gsa}. 
The value of the cross section of coherent $\rho$ meson photoproduction on lead 
at the photon--nucleon center-of-mass energy of $W_{\gamma p}\approx 46$ GeV   
extracted from this measurement is $\sigma_{\gamma Pb\rightarrow \rho Pb}\approx 2$ mb,
which is very close to the $\gamma Au\rightarrow \rho Au$ cross sections 
in the range of energies of $W_{\gamma p}\approx 7 - 12 $ GeV obtained by the
STAR collaboration in ion--ion UPCs at RHIC~\cite{Adler:2002sc,Abelev:2007nb,Agakishiev:2011me}.
While the calculations of the standard elastic nuclear shadowing in the Glauber model capture
the bulk of the nuclear suppression by the factor of approximately four, 
the ALICE
experimental values are still significantly, by the factor of approximately  $1.5 - 1.7$, smaller than  the VMD-GM
predictions of~\cite{Frankfurt:2002wc,Frankfurt:2002sv,Rebyakova:2011vf}. 
It is also important to emphasize that the VMD model with the standard values of the photon--$\rho$ meson coupling
constant $f_{\rho}$ and the $\rho$--nucleon cross section $\sigma_{\rho N}$ overestimates the most recent H1 
data on diffractive $\rho$ photoproduction on the proton at HERA~\cite{Weber:2006di} by the factor of approximately $1.3$.

The aim of the present paper is to modify both VMD and Glauber models 
at high photon energies to take into account
an increasing role of hadronic fluctuations of the photon
interacting with different strength and having a wide range of masses.

At high energies, the Glauber model 
is substituted by the Gribov--Glauber approach~\cite{Gribov:1968jf},
which takes into account both elastic and 
inelastic diffraction in the intermediate states contributing to the shadowing correction.
Note that in spite of very different pictures of the space--time evolution of the scattering process
at moderate and high energies, the expressions for the shadowing correction 
to hadron--nucleus cross sections look similar in the two approaches.
In particular, 
the nuclear shadowing is calculable in terms of the elementary hadron--nucleon diffractive 
cross section, which includes the elastic hadron--nucleon scattering 
leading to the elastic nuclear shadowing correction of the Glauber model
and the inelastic hadron--nucleon scattering.
With an increase of energy, inelastic diffraction becomes essential and this
results in Gribov's inelastic nuclear shadowing (GINS) correction.

While an increase of the GINS correction with energy is well-known in the discussed energy range, it is not often 
appreciated that the relative magnitude of the effect is larger for projectiles with smaller hadron--nucleon 
cross sections.  
Indeed, in proton--nucleus scattering, 
the GINS correction to the total proton--nucleus cross sections is found to be 
small~\cite{Karmanov:1973va,CiofidegliAtti:2011fh}.
As a consequence, the Glauber model is now widely used in the analysis and interpretation of the data on 
high energy heavy ion collisions at RHIC and the LHC. For pion--nucleus scattering, the analysis of total 
pion--deuteron~\cite{Dakhno:1982ez} and pion--nucleus~\cite{Dersch:1999zg}
cross sections demonstrated that the inelastic nuclear shadowing correction is larger than
that in the proton-nucleus interaction.
We argue that in the high energy $\rho$ meson photoproduction on nuclear targets the GINS correction
can be even stronger than in the pion--nucleus case.

We pointed out
above that an increase of the photon inelastic diffraction into large masses 
can be understood as resulting from photon fluctuations with the large invariant masses.
These fluctuations lead to a certain decrease of the 
$\gamma N \to \rho N$ cross section compared to the expectations based on the VMD and the additive quark 
model assumption $\sigma_{\rho N}=\sigma_{\pi N}$.
Indeed, in the GVMD model, the partial cancellation between the diagonal $\rho N \to \rho N$ and
the non-diagonal $\rho^{\prime} N \to \rho N$ transitions, which is required to restore the 
Bjorken scaling of the total virtual photon--proton cross section, naturally leads to a decrease of 
$\sigma_{\gamma N \to \rho N}$. In the CDM, the quark--antiquark dipoles with the large relative transverse momentum
$k_t$ and the longitudinal momentum fraction $z \neq 0,1$ are characterized by the large invariant mass, 
the small transverse size $d_t$ and, hence, the
reduced cross section due to color transparency of QCD. Since the overlap integral between the photon and $\rho$ meson wave functions has a stronger support at small dipole sizes $d_t$ than the square the $\rho$ meson (pion) wave function due to the 
point-like photon coupling to quarks, the contribution of such dipoles somewhat decreases $\sigma_{\gamma N \to \rho N}$.
In the framework of the VMD model, this reduction can be explained either by an increase of $f_{\rho}$ or a decrease of 
$\sigma_{\rho N}$. 
Since the former is very well constrained by the measured $\rho \to e^{+} e^{-}$ decay width, 
one concludes that $\sigma_{\rho N}$ should be somewhat reduced, $\sigma_{\rho N} < \sigma_{\pi N}$.

This results in an overall decrease of the $\gamma N \to \rho N$ cross section and
the relative increase of the GINS correction.
Besides, the effect of the inelastic shadowing correction 
in the elastic cross section is larger than that in the total cross section. 
Hence, these
modifications of the VMD-GM approach   
noticeably reduce the cross section of coherent $\rho$ photoproduction on nuclei.

To implement these effects in our calculations, we (i) use the framework of cross section 
fluctuations~\cite{Good:1960ba, Miettinen:1978jb, Kopeliovich:1981pz, Blaettel:1993ah, Frankfurt:1994hf,Harrington:1995qb} 
taking into account  quark counting rules for the probability of cross section fluctuations for small $\sigma$ 
and the information on inelastic diffraction in 
photon scattering off a nucleon target, and (ii) modify the
VMD model to effectively include the effect of the reduction of the $\rho$--nucleon cross section.
The resulting approach allows us to 
describe well 
the data on elastic $\rho$ photoproduction on nuclei in heavy ion 
UPCs in the 
$7 \ {\rm GeV} < W_{\gamma p} < 46$ GeV 
energy range in a way consistent with the 
$\gamma p\rightarrow \rho p$ HERA 2006 data \cite{Weber:2006di}, 
and, thus, to explain away the discrepancy between the data on the
$\gamma A \to \rho A$ cross section and its theoretical description in the VMD-GM approach.

\section{Cross section of $\rho$ photoproduction off nuclear targets  from STAR and ALICE UPC measurements}
\label{sec:excs}

The basic expression for the cross section of vector meson $V$ photoproduction in 
nucleus--nucleus UPCs (for review of the physics of ultraperipheral collisions 
and references, see \cite{Baltz:2007kq}) reads:
\begin{equation}
 \frac{d \sigma_{A A\to AA V }(y)}{dy} 
=N_{\gamma/A}(y)\sigma_{\gamma A\to V A}(y)+
N_{\gamma/A}(-y)\sigma_{\gamma A\to V A}(-y) \,,
\label{csupc}
\end{equation}
where $y = \ln(2\omega/M_{V})$ is the rapidity of the vector meson;   $\omega$ is the photon energy; 
$M_V$ is the vector meson mass;  $\sigma_{\gamma A\to V A}$ is the cross section of exclusive 
coherent photoproduction of $V$ on nucleus $A$; $N_{\gamma/A}(y)$ is the photon flux. 
The presence of two terms in Eq.~(\ref{csupc}) is the reflection of 
the fact that each colliding nucleus can serve
as a photon source and as a target.  

The flux of photons produced by a fast-moving 
point-like charge is well-known from classical electrodynamics:
\begin{equation}
N_{\gamma/A}(y)=\frac{2 Z^2 \alpha_{\rm e.m.}}
{\pi}\left[\zeta K_0(\zeta) K_1(\zeta)-\frac{\zeta^2}{2}\left(K_1^2(\zeta)-K_0^2(\zeta)\right)  \right] \,,
\label{eq:flux}
\end{equation}
where $Z$ is the nucleus charge; 
$\alpha_{\rm e.m.}$ is the fine-structure constant; $K_{0}$
and $K_1$ are modified Bessel functions of the
second kind; $\zeta=M_{V}b_{\rm min}e^{-y} /(2\gamma_L)$, 
where $\gamma_L$ is the nucleus Lorentz factor and $b_{\rm min}$ is
the minimal impact parameter of the nucleus--nucleus ultraperipheral collision.
The photon flux calculated using Eq.~(\ref{eq:flux}) 
with $b_{\rm min}=2 R_A$ ($R_A$ is the nuclear radius)
reproduces with the precision of a few percent a more accurate calculation, which 
takes into account the nuclear form factor and the suppression of the strong
nucleus--nucleus interaction calculated using the Glauber model,
see the discussion and references in~\cite{Guzey:2013taa,ALICE:2012aa}. In the latter work,  
the very good precision of the theoretical calculation of $N_{\gamma/A}(y)$ was explicitly demonstrated by 
the measurement of neutron emission in electromagnetic dissociation of Pb nuclei at the
LHC at $\sqrt{s_{NN}}=2.76$ TeV.

At $y=0$, the two terms in Eq.~(\ref{csupc}) are equal and one 
can then unambiguously determine the
$\gamma A \to \rho A$ cross section using the experimental 
values of the $d \sigma_{A A\to AA \rho}(y=0)/dy$
cross section measured by the STAR collaboration at 
RHIC~\cite{Adler:2002sc,Abelev:2007nb,Agakishiev:2011me} 
and the ALICE collaboration at the LHC~\cite{Adam:2015gsa}:
\begin{equation}
\sigma_{\gamma A \to \rho A}(W_{\gamma p})=\frac {1} {2N_{\gamma/A}(y=0)}
\frac{d \sigma_{AA\to AA \rho }(y=0)}{dy} \,,
\label{csgama}
\end{equation}
where 
$W_{\gamma p} \equiv W_{\gamma p}(y=0)=\sqrt{2M_{\rho}m_{N}\gamma_{L}}$ 
with $M_{\rho}=770$ MeV being the $\rho$ meson mass and 
$m_N$ the nucleon mass.
The $\gamma A \to \rho A$ cross section determined this way is presented in
Table~\ref{table:ecstable} as a function of the corresponding $W_{\gamma p}$.
Note that similarly to the invariant energy of $AA$ collisions per nucleon $\sqrt{s_{NN}}$,  
throughout our paper $W_{\gamma p}$ denotes the invariant photon--nucleus energy per nucleon.

 \begin{table}[htbp]
    \begin{center}
      \begin{tabular}
	{|c|c|c|c|}  \hline
 Nuclear target & $ W_{\gamma p}$, GeV & $\sigma_{\gamma A \to \rho A}$, mb  \\ \hline
 $\gamma Au\rightarrow \rho Au$ & 6.96  &  $2.08 \pm 0.33$\\ \hline
 $\gamma Au\rightarrow \rho Au$ & 10.04  & $ 1.9 \pm 0.76$   \\ \hline
 $\gamma Au\rightarrow \rho Au$ & 12.46  &  $1.58 \pm 0.25$ \\ \hline
 $\gamma Pb\rightarrow \rho Pb$ & 46.28 & $1.97\pm 0.21$    \\ \hline
      \end{tabular}      
\caption{The cross sections of $\rho$ photoproduction on nuclear  
targets extracted from the STAR
~\cite{Adler:2002sc,Abelev:2007nb,Agakishiev:2011me} and 
the ALICE UPC measurements~\cite{Adam:2015gsa}.}
      \label{table:ecstable}
    \end{center}
  \end{table}

\section{Nuclear shadowing in $\rho$ photoproduction on nuclear targets}
\label{sec:Shadowing}

In the Glauber model, the cross section of coherent $\rho$ photoproduction on nuclei reads~\cite{Bauer:1977iq}:
\begin{equation}
\sigma_{\gamma A \to \rho A}^{\rm VMD}=
\frac{d\sigma_{\gamma p \to \rho p}(t=0)}{dt}
 \int_{-\infty}^{t_{\rm min}} dt 
\left|\int d^2 \vec{b} \,
e^{i \vec{q}_{\perp} \cdot \vec{b}} \int dz \rho_A (b,z) e^{i q_{\parallel} z} 
e^{-(1-i\eta) \frac{\sigma_{\rho N}}{2} \int_z^{\infty} dz^{\prime} 
\rho_A(b,z^{\prime})}
 \right|^2 \,,
 \label{eq:cs_rho}
\end{equation}
where 
$\sigma_{\rho N}$ is the total $\rho$--nucleon cross section; 
$\eta$ is the ratio of the real to the imaginary parts
of the $\rho$--nucleon scattering amplitude;
$\vec{q}_{\perp}$ and $q_{\parallel}$ are the transverse and 
longitudinal components of the momentum transfer to the nucleus, respectively;
$\rho_A$ is the nuclear density normalized by the relation 
$\int dz\,d^2 \vec{b}\rho_A (b,z)=A$. The nuclear density is
well known from studies of elastic electron and proton scattering on
nuclei at intermediate energies (see, for example, 
\cite{Friar:1973wy,Friar:1975pp,Alkhazov:1978et}).
In our calculations we used
the Hartree--Fock--Skyrme nuclear density
which describes the root-mean-square radii of nuclei 
across the periodic table with the
precision better than $2\%$~\cite{Beiner:1974gc}. The minimal momentum transferred squared is 
$t_{\rm min}=-(M_{\rho}^2 m_N/W_{\gamma p}^2)^2=-(q_{\parallel})^2$,
where 
 $W_{\gamma p}$ is the 
invariant photon--nucleon energy ($W_{\gamma p}^2=2 m_N E_{\gamma}+m_N^2$ 
in the laboratory frame).
In the VMD model, the forward elementary $\gamma p \to \rho p$ cross section
in Eq.~(\ref{eq:cs_rho}) can be related to the total $\rho N$ cross section,
$\sigma_{\rho N}$, using the optical theorem:
\begin{equation}
\frac{d\sigma_{\gamma p \to \rho p}(t=0)}{dt}=
\frac{1}{16 \pi} \left(\frac{e}{f_{\rho}}\right)^2
 (1+\eta^2) \sigma_{\rho N}^2 \,,
\label{eq:sigma_elem}
\end{equation}
where $f_{\rho}$ is the $\gamma - \rho$ coupling constant
($f^2_{\rho}/4\pi =2.01\pm 0.1$) fixed by 
the $\Gamma (\rho\rightarrow e^+ e^- )$ width of the $\rho \to e^{+} e^{-}$ decay.
Note also that the effect of $t_{\rm min} \neq 0$ in  
$d\sigma_{\gamma p \to \rho p} (t=0)/dt$
 is negligibly small 
and can be neglected.

At high photon energies, neglecting the effects of $q_{\parallel} \neq 0$ 
and $\eta \neq 0$ in 
Eq.~(\ref{eq:cs_rho}) (in the $5\,{\rm GeV} <W_{\gamma p}<50\, {\rm GeV}$ range, 
$|\eta| <0.1$ and thus can be safely neglected~\cite{Donnachie:1992ny}),
one can  write it in the 
form of the optical limit of 
the Glauber model:
\begin{equation}
\sigma_{\gamma A \to \rho A}^{\rm VMD}=
\left(\frac{e}{f_{\rho}}\right)^2  \int d^2\vec{b} 
\left|1-e^{-\frac{\sigma_{\rho N}}{2}  T_A(b)}\right|^2 =
\left(\frac{e}{f_{\rho}}\right)^2 \sigma^{\rm el}_{\rho A} \,,
\label{eq:cs_rho_approx}
\end{equation}
where $T_A(b)=\int dz \rho_A(b,z)$.

The interpretation of Eq.~(\ref{eq:cs_rho_approx}) is 
straightforward and well-known: the incoming photon 
transforms into a $\rho$ meson far before the target and
then the
$\rho$ meson
coherently interacts with the 
nucleons along its trajectory. 
The Glauber model takes into account  the elastic
nuclear shadowing effect, which depends on the total $\rho N$ 
cross section.  Based on the additive
quark 
model, it is generally assumed 
that 
$\sigma_{\rho N}(W)=\sigma_{\pi N}(W)=[\sigma_{\pi^+ N}(W)+\sigma_{\pi^- N}(W)]/2$,
where $W$ is the invariant $\rho$ meson (pion)--nucleon energy.
The experimental pion-nucleon cross sections in a wide range of energies
are well described 
as a simple sum of the soft Pomeron and the secondary Reggeon exchanges~\cite{Donnachie:1992ny}.
Correspondingly, this resulted in a simple form (DL94) for 
the energy  dependence of $\sigma_{\rho N}$ \cite{Donnachie:1994zb}:
\begin{equation}
\sigma_{\rho N}(W)= 
13.6 \, W^{2(\alpha_{P}(0)-1)}+31.8 \, W^{2(\alpha_R (0)-1)} \,,
\label{pics}
\end{equation}
where $\alpha_R (t)=0.55+0.93 t$ is the Reggeon exchange trajectory;
$\alpha_P (t)=1.0808+0.25t$ is the soft 
Pomeron trajectory
characterizing
the high energy behavior of soft 
hadron--nucleon processes. 
Since
the DL94 model did not fit well the forward
$\rho$ photoproduction cross section on the proton 
measured to that time
in low energy experiments~\cite{Park:1971ts,Aston:1982hr,Egloff:1979mg} and in the ZEUS 
experiment~\cite{Derrick:1995vq,Derrick:1996vw,Breitweg:1997ed} at high
energies, 
Donnachie and 
Landshoff suggested~\cite{Donnachie:1999yb} to  
simply renormalize the 
forward cross section [Eq.~(\ref{eq:sigma_elem})] by the factor of 0.84 
motivated by possible corrections to the 
$\gamma-\rho$  coupling constant 
(this renormalization was consistent with large experimental uncertainties  of the data).
\begin{figure}[h]
\begin{center}
\epsfig{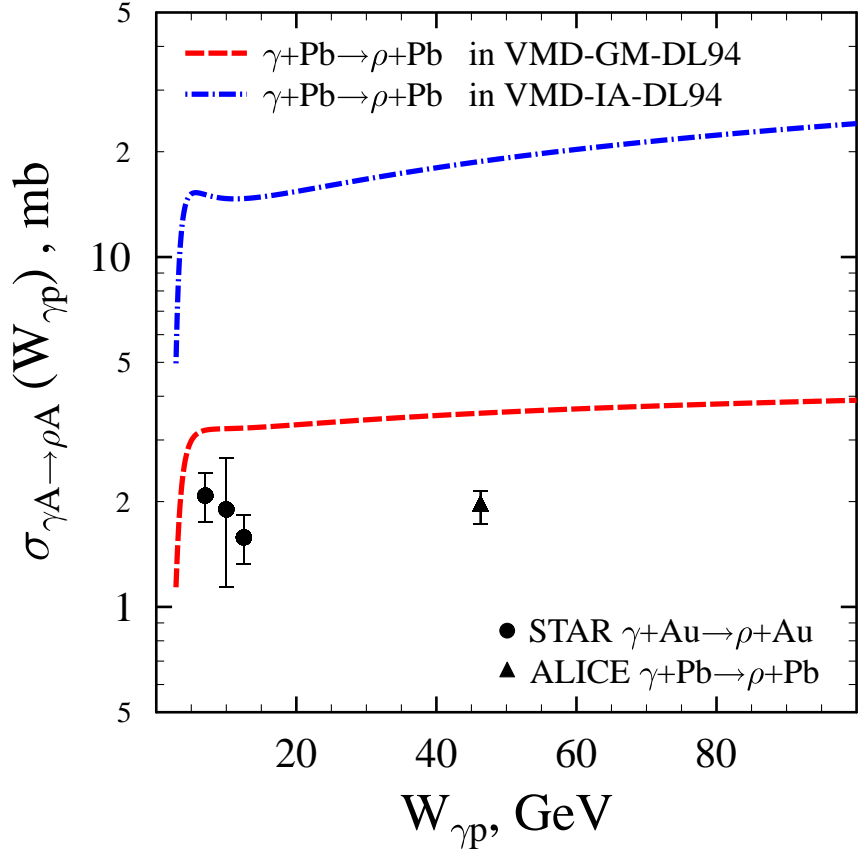}
 \caption{
 The $\gamma A \to \rho A$ cross section as a function of $W_{\gamma p}$.
 The 
 VMD-GM (red dashed curve) and VMD-IA (blue dot-dashed line)
 predictions for a $^{208}$Pb target based on the DL94 
parametrization of the
$\rho N$ cross section 
are compared to the experimental values extracted 
from the STAR and ALICE UPC measurements.}
 \label{fig:csgimp}
\end{center}
\end{figure}

In Fig.~\ref{fig:csgimp} we compare the total cross sections 
of coherent $\rho$ photoproduction
extracted from the STAR and ALICE measurements 
(see Table~\ref{table:ecstable})
to those calculated 
in the impulse approximation (IA), when all nuclear effects except for
coherence are neglected (blue dot-dashed line),
and in the Glauber model (red dashed line) using the  DL94 $\rho N$ total cross section (Eq.~(\ref{pics})).  
From the figure one finds that 
the VMD-GM with the DL94 model for $\sigma_{\rho N}$ 
predicts the suppression of $\sigma_{\gamma A \to \rho A}$
by approximately a factor of four compared to the IA calculation,
but it still
overestimates the experimental cross sections  by the factor of $1.5 - 2$.
Besides, the energy dependence is different: while the calculated cross sections 
slowly grow with energy, 
the experimental values 
slightly decrease or stay almost constant. 
 Note that 
 the calculated values of the $\gamma Au\rightarrow \rho Au$
cross section 
are smaller than those for the lead target by approximately 5\% for all energies.
Hence, we neglect this difference throughout our paper and perform our calculations for lead keeping
in mind the 5\% reduction of the nuclear cross section when we compare our calculations
with the STAR data.
\begin{figure}
\begin{center}
\epsfig{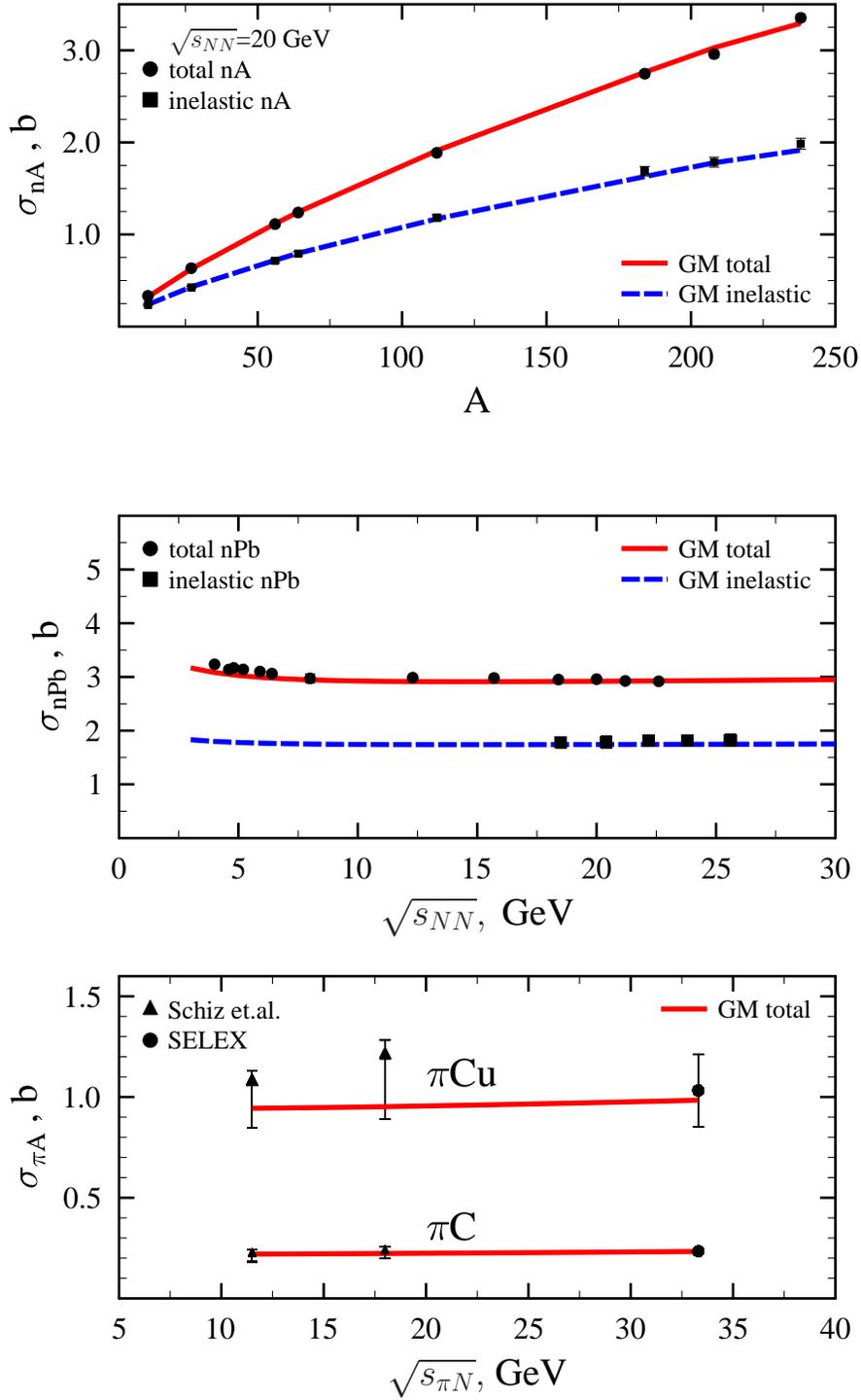}
 \caption{Upper
  and middle:  Comparison of the  total and inelastic neutron--nucleus cross sections calculated in the Glauber model with the available data.
Bottom: The total pion--nucleus cross section as a function of 
$\sqrt{s_{\pi N}}$:
the Glauber model calculations with the DL94 model for $\sigma_{\pi N}$ are compared to the available data.}
 \label{fig:csha}
\end{center}
\end{figure}

To check  
the accuracy of the Glauber model calculations in
Eq.~(\ref{eq:cs_rho_approx}) in combination with the DL94 pion--nucleon
cross section,
we calculated the 
hadron--nucleus total and inelastic cross sections for the neutron and pion projectiles
in the Glauber approach:
\begin{eqnarray}
\sigma^{\rm tot}_{hA} &=& 2\int d^2 {\vec b}
\left [1-e^{-\frac{\sigma_{h N}}{2}  T_A(b)}\right ] \,, \nonumber\\
\sigma^{\rm in}_{hA} &=& \int d^2 {\vec b}
\left [1-e^{-{\sigma_{h N}}  T_A(b)}\right ] \,.
\label{haincs}
\end{eqnarray} 
The neutron--nucleon cross section $\sigma_{nN}$ 
is estimated using the additive quark model counting rule
relation~\cite{Levin:1965mi}
$\sigma_{nN}=3/2  \sigma_{\pi N}$, where
the pion--nucleon cross section is given by Eq.~(\ref{pics}). 
The results of our calculations are compared to the data~\cite{Dersch:1999zg,Ramana Murthy:1975qb,Roberts:1979jf,Schiz:1979qf}
in Fig.~\ref{fig:csha}. 
One can see from the figure that the calculations agree very well
with the measurements. This 
means that
the reasons of the disagreement 
of similar calculations of the $\gamma A\rightarrow \rho A$ cross section 
with the STAR and ALICE data
are in specifics of the light vector meson photoproduction process.

\begin{figure}
\begin{center}
\epsfig{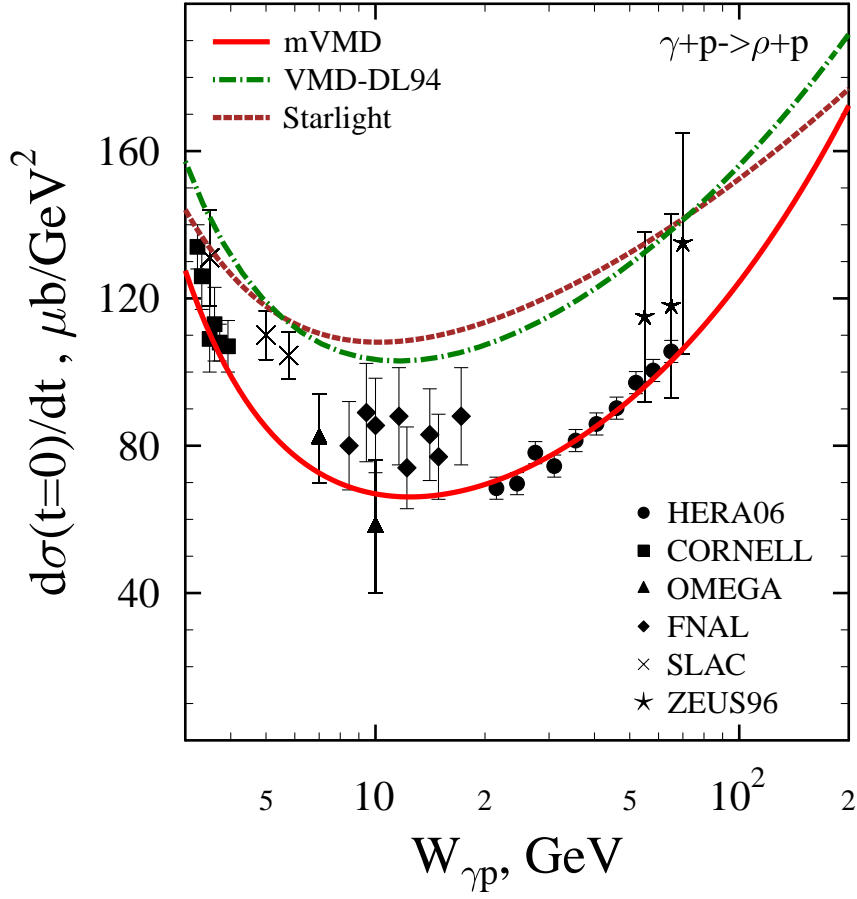}
 \caption{Comparison of the experimentally measured forward
 cross section of coherent $\rho$ photoproduction on the proton~\cite{Weber:2006di,Park:1971ts,Aston:1982hr,Egloff:1979mg,
{Derrick:1995vq},{Derrick:1996vw},Breitweg:1997ed}
with the
 VDM-DL94 model and the Starlight parametrization. The red solid line
 shows the modified VMD (mVMD) parametrization 
 (see text for details).}
 \label{fig:freecs}
\end{center}
\end{figure}

This conclusion is confirmed by our 
observation that 
the latest 2006 H1 data on the $\gamma p\rightarrow  \rho p$ 
cross section~\cite{Weber:2006di}
(we extrapolated the H1 cross sections 
given at $-t=0.01$ GeV$^2$ to $-t=0$ assuming the $e^{Bt}$ dependence with
the value of the slope $B$ reported by H1)
disagrees with the normalization of the forward cross section calculated using the DL94 model by the factor of 0.84.
This is seen in Fig.~\ref{fig:freecs}, where
the forward $\gamma p\rightarrow \rho p$ cross section evaluated using
Eqs.~(\ref{eq:sigma_elem}) and (\ref{pics}) 
(the green dot-dashed curve labelled ``VMD-DL94")
is compared to the whole bulk of the data. 
Also, for comparison, we show the parametrization of the 
forward $\gamma p\rightarrow \rho p$ cross section 
from the Starlight 
Monte-Carlo generator~\cite{Klein:1999qj}, which is widely used for 
predictions and modeling 
of vector meson photoproduction on nuclear targets. 
In order to agree with the 2006 H1 data, 
the results of the VMD-DL94 and the Starlight parametrization 
should be decreased by the factor of approximately
$0.7$,
which is much larger than what could be 
allowed by a variation of
$f_{\rho}$.   
From the analysis presented above we can 
conclude the following: 
the assumption of the
$\rho$ meson dominance in the photon wave function
has to be modified in order to agree to the
whole set of data including the results of 2006 H1 measurements.

To this end, 
one can write the $\rho$ meson photoproduction amplitude  as the dispersion 
integral over the masses of the intermediate states generated in 
the $\gamma \to V$ transitions, 
which will involve the on-mass-shell $f_{V}$, the $\rho N$ cross section
and the $VN \to \rho N$ amplitude 
(here $V$ denotes $\rho$-meson-like fluctuations of the photon with the invariant mass $M$, see our discussion
in the Introduction).
It is possible to demonstrate that inclusion of the contribution of 
the higher states can only weakly change $f_{\rho}$, but 
it can noticeably reduce the cross 
section of the $\rho$ meson production due non-diagonal transitions among different hadronic components 
of the  photon and the $\rho$ meson in the GVMD approach~\cite{Fraas:1974gh,Ditsas:1976yv,Pautz:1997eh}.
On the other hand, within the VMD approach this can be modeled by defining the  
effective $\rho$--nucleon cross section $\hat{\sigma}_{\rho N}$:
\begin{eqnarray}
\hat{\sigma}_{\rho N} 
(W_{\gamma p})={\frac {f_{\rho}} {e}}\sqrt{{16 \pi}
\frac{d\sigma^{\rm exp}_{\gamma p \to \rho p}(t=0)}{dt}} \,.
\label{eq:sigma_hat}
\end{eqnarray}
We refer to this model 
as the modified vector meson dominance (mVMD) model; 
its prediction is shown by the solid red curve in Fig.~\ref{fig:freecs}.
Note that a similar 
effect is also present in the CDM.

The Gribov--Glauber model takes into account both elastic and 
inelastic diffraction; 
the latter leads to the additional---as compared to the Glauber model---inelastic 
nuclear shadowing
contribution
(the Gribov shadowing 
correction)~\cite{Gribov:1968jf}.
The standard method to include this effect
is given by the formalism of cross section fluctuations,
which conveniently and successfully describes diffractive 
dissociation of protons, neutrons and pions on hydrogen 
and nuclei and inelastic nuclear shadowing in hadron--nucleus 
total cross sections~\cite{Frankfurt:2000tya}.

Applying this formalism to the $\rho$ meson--nucleus scattering,
we obtain:
\begin{equation}
\sigma_{\gamma A \to \rho A}^{\rm mVMD-GGM}=\left(\frac{e}{f_{\rho}}\right)^2  
\int d^2\vec{b} \left| \int d\sigma P(\sigma)
\left(1-e^{-\frac{\sigma}{2}  T_A(b)} \right)\right|^2  \,,
\label{eq:cs_rho_approx_CF}
\end{equation}
which generalizes Eq.~(\ref{eq:cs_rho_approx}).

The interpretation of Eq.~(\ref{eq:cs_rho_approx_CF}) 
is the following: the photon fluctuates into the $\rho$ meson,
which interacts with the target as a coherent superposition of 
eigenstates of the scattering operator, whose eigenvalues are 
the scattering cross sections $\sigma$;
the weight of a given fluctuation is given by the distribution $P(\sigma)$.
Each state interacts with nucleons of the target nucleus 
according to the Gribov--Glauber model.
The result is summed over all possible fluctuations,  
which corresponds to averaging with the distribution 
$P(\sigma)$ at the amplitude level.

Based on the similarity between the pion and $\rho$ meson wave functions suggested by 
the additive quark model and our discussion above,
it is natural to assume that
$P(\sigma)$ for the $\rho N$ interaction 
should be similar to the pion $P_{\pi}(\sigma)$, 
which  we additionally
multiply by the factor of $1/(1+(\sigma/\sigma_0)^2)$ to
take into account the enhanced contribution
of small $\sigma$ in the $\rho N$ interaction
(we explained above  that the contribution of small-$\sigma$ fluctuations to the $\gamma N \to \rho N$
amplitude is expected to be enhanced compared to the $\pi N \to \pi N$ one): 
\begin{equation}
P(\sigma)=C \frac{1}{1+(\sigma/\sigma_0)^2} e^{-(\sigma/\sigma_0 -1)^2/\Omega^2} \,.
\label{eq:Psigma}
\end{equation}
The parameterization of Eq.~(\ref{eq:Psigma}) satisfies the basic QCD constraint of $P(\sigma=0)\neq 0$ and
also $P(\sigma \rightarrow \infty)\rightarrow 0$.
The free parameters $C$, $\sigma_0$ and $\Omega$ are found from 
the following constraints:
\begin{eqnarray}
\int d\sigma P(\sigma) &=& 1 \,, \nonumber\\
\int d\sigma P(\sigma) \sigma  &=&  \langle \sigma \rangle\,, \nonumber\\
\int d\sigma P(\sigma) \sigma^2  &=& \langle \sigma \rangle^2 (1+\omega_{\sigma}) \,,
\label{eq:P_pion2}
\end{eqnarray}
where $\langle \sigma \rangle = \hat{\sigma}_{\rho N}$ in the mVMD model,
see Eq.~(\ref{eq:sigma_hat}).  

The quantity $\omega_{\sigma}$ parametrizes 
the dispersion of $P(\sigma)$ around its mean value $\langle \sigma \rangle$, i.e.,
it characterizes the strength of cross section fluctuations.
It can be determined using experimental information on the photon 
diffraction dissociation, in particular,
the factorization of the photon and the pion diffraction dissociation 
cross sections scaled by the 
respective total cross sections. In detail, 
the measurement~\cite{Chapin:1985mf} of inclusive diffraction dissociation 
of photons on hydrogen, $\gamma p \to X p$, 
in the range of
$75 < E_{\gamma} < 148$ GeV and $M_X^2/s < 0.1$ ($M_X$ denotes 
the produced diffractive mass) and 
the control measurement of inclusive diffraction dissociation 
of pions in the $\pi p \to X p$ reaction at 
$E_{\pi}=100$ GeV showed that the respective $M_X^2$ distributions 
scaled by the total cross sections are very similar in the 
photon and pion cases. For the cross sections integrated over $M_X^2$, 
this observation means that: 
\begin{equation}
\frac{d\sigma_{\gamma p \to Xp}(t=0)/dt}{\sigma_{\gamma p}} 
\approx \frac{d\sigma_{\pi p \to Xp}(t=0)/dt}{\sigma_{\pi p}}
=\frac{\omega_{\sigma}^{\pi}}{16 \pi} \sigma_{\pi N} \,,
\label{eq:Chapin}
\end{equation}
where in the last equation we expressed the cross section of pion 
diffraction dissociation in terms of 
$\omega_{\sigma}^{\pi}$ 
characterizing the $P_{\pi}(\sigma)$ distribution and the total 
pion--nucleon cross section $\sigma_{\pi N}$.

On the other hand, using the formalism of cross section fluctuations 
for the $\rho$--nucleon scattering and the 
mVMD model for the $\gamma-\rho$ transition, we obtain 
for the cross section of photon diffraction dissociation
[compare to Eq.~(\ref{eq:sigma_elem})]:
\begin{equation}
\frac{d\sigma_{\gamma p \to X p}(t=0)}{dt}=
\frac{1}{16 \pi} \left(\frac{e}{f_{\rho}}\right)^2 
\left[\int d\sigma P(\sigma) \sigma^2-(\hat{\sigma}_{\rho N})^2\right]=
\frac{\omega_{\sigma}}{16 \pi} \left(\frac{e}{f_{\rho}}\right)^2  
(\hat{\sigma}_{\rho N})^2\,,
\label{eq:sigma_elem_CF}
\end{equation}
where the diffraction dissociation final state $X$ by construction 
does not contain $\rho$. 
The inelastic final state $X$ is selected experimentally by analyzing the differential cross section as a function of the 
produced diffractive mass $M_X$ and corresponds to the values of $M_X$ beyond the $\rho$ peak, $M_X^2 > 1.5-2$ GeV$^2$~\cite{Chapin:1985mf}.
Substituting 
Eq.~(\ref{eq:sigma_elem_CF}) in Eq.~(\ref{eq:Chapin}) 
we obtain the desired constraint on $\omega_{\sigma}$:
\begin{equation}
\omega_{\sigma}=\frac {f_{\rho}^2} {e^2}
 \frac{\sigma_{\pi N}\sigma_{\gamma p}}
{\hat{\sigma}^2_{\rho N}} \omega_{\sigma}^{\pi} \,,
\label{eq:omega_VMD}
\end{equation}
where the total photon--proton cross section $\sigma_{\gamma p}$ is taken from the fit to data~\cite{Donnachie:1992ny}.

For the pion projectile, we use the constituent quark counting 
rule for the ratio of the nucleon--nucleon and the 
pion--nucleon total cross sections and obtain:
\begin{equation}
\omega^{\pi}_{\sigma}(s)=\frac{3}{2}\, \omega^{N}_{\sigma}\left(s\right) \,.
\label{eq:omega_fit_pion}
\end{equation}
Here we effectively use validity of the limiting fragmentation 
which is well established experimentally.

The pattern of cross section fluctuations for the nucleon projectile has 
the following dependence of the 
invariant collision energy $\sqrt{s}$:
the cross section fluctuations reach a broad maximum for $24 < \sqrt{s}< 200$ GeV, 
are most likely 
small for $\sqrt{s} < 24$ GeV and gradually decrease for $\sqrt{s} > 200$ GeV 
toward the Tevatron and LHC energies.
Therefore, we use the following parametrization for the 
parameter $\omega_{\sigma}^N$ describing the dispersion of the fluctuations:
\begin{equation}
\omega^N_{\sigma}(s)=\left\{\begin{array}{ll}
\beta \,\sqrt{s}/24 \,, & \sqrt{s} < 24 \ {\rm GeV} \,, \\
\beta  \,, &  24 < \sqrt{s} < 200 \  {\rm GeV} \,, \\
\beta  -0.15 \ln (\sqrt{s}/200)+0.03 (\ln (\sqrt{s}/200))^2 \,, &  \sqrt{s} > 200 \ {\rm GeV} \,.
\end{array} \right.
\label{eq:omega_fit}
\end{equation}
where the parameter $\beta \approx 0.25 - 0.35$ was determined from the 
analysis of $pp$ and $\bar p p$ data \cite{Blaettel:1993ah}.

\begin{figure}[ht]
\begin{center}
\epsfig{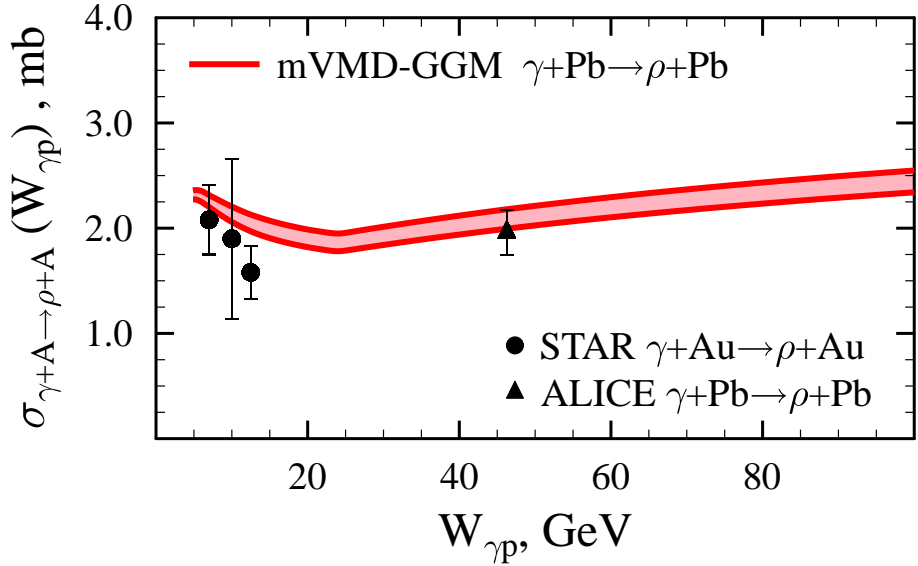}
 \caption{The $\sigma_{\gamma A \to \rho A}$  cross section as a 
function of $W_{\gamma p}$. 
The theoretical predictions using the mVMD model for the $\gamma p \to \rho p$ cross section and the 
Gribov--Glauber model with cross section fluctuations for the $\gamma A \to \rho A$ amplitude are compared to the 
STAR (circle) and ALICE (triangle) data.
The shaded area reflects the theoretical uncertainty associated 
with the parameter $\beta$ characterizing
the strength of cross section fluctuations
(see text for details).}
 \label{fig:sigma_final_wf_CF_new}
\end{center}
\end{figure}

It is known \cite{CiofidegliAtti:2011fh}
from studies of corrections to the Glauber model 
for total proton--nucleus cross sections that suppression due to the inelastic
shadowing is almost compensated by 
the effect of 
short-range correlations (SRC)
in the wave function of the target nucleus. We included the effect of SRC 
by the following replacement~\cite{Moniz:1971ih}:
\begin{equation}
T_A(b) \to T_A(b)+\xi _c \frac{\sigma_{\rho N}}{2} \int dz \rho_A^2(b,z) \,,
\label{eq:correlations}
\end{equation}
where $\xi _c=0.74$ fm is the correlation length.

Our predictions
 for the $\gamma A \to \rho A$ cross section as a function of $W_{\gamma p}$
 are presented in Fig.~\ref{fig:sigma_final_wf_CF_new}. 
The 
shaded area spanned by two red curves
presents the results of the calculation using the
 mVMD model for the $\gamma p \to \rho p$ cross section and the 
 Gribov--Glauber model with the effect of cross section 
 fluctuations, see Eq.~(\ref{eq:cs_rho_approx_CF}).  
The shaded area shows the uncertainty
of our calculations 
due to the variation of the fluctuation strength $\omega_{\sigma}$
by changing 
$\beta$ in the range $0.25\leq \beta \leq 0.35$.
Our predictions are 
compared to the 
STAR (circle) and ALICE (triangle) data.
One can clearly see from the figure that the inclusion of 
the inelastic nuclear shadowing enables us to explain the discrepancy 
between the UPC data on coherent $\rho$ photoproduction on nuclei 
at large $W_{\gamma p}$ and the theoretical description of 
this process in the framework of the VMD-GM with the DL94  parametrization
of the $\rho N$ cross section.

\section{Discussion}
\label{sec:discussion}

\begin{figure}[t]
\begin{center}
\epsfig{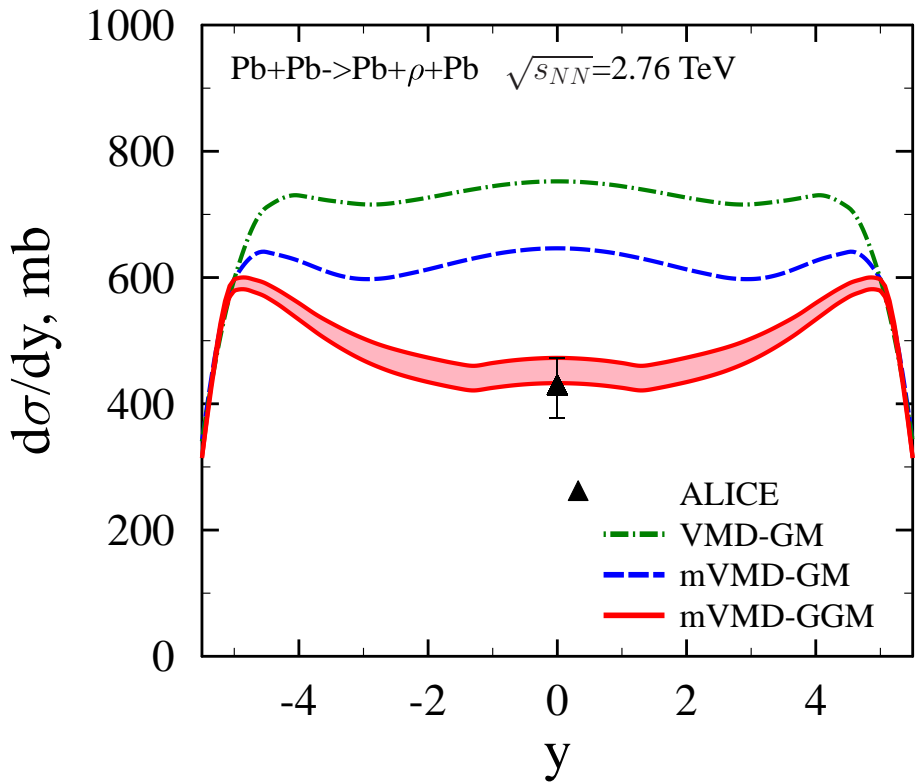}
 \caption{The rapidity distribution
 of coherent $\rho$ photoproduction
in Pb-Pb UPCs at $\sqrt{s_{NN}}=2.76$ TeV.
Theoretical predictions of the mVDM-GGM (red solid curves with
the shaded area showing the uncertainty due to the variation of the fluctuation strength),
the mVMD-GM (blue dashed curve) and the VMD-GM (green dot-dashed curve) are compared to the ALICE data
(see text for details).
} 
 \label{fig:rad_dist}
\end{center}
\end{figure}

The effect of the inelastic shadowing correction, which we demonstrate
in these calculations, can be checked in the UPC
measurements at the LHC.
The inelastic nuclear shadowing changes the rapidity 
distribution of coherent $\rho$ photoproduction
in ion UPCs. Figure~\ref{fig:rad_dist} presents the results of our 
calculation of $d\sigma_{Pb Pb \to \rho Pb Pb}/dy$,
see Eq.~(\ref{csupc}),
as a function of the $\rho$ meson rapidity $y$ in Pb-Pb UPCs at 
the LHC at 
$\sqrt{s_{NN}}=2.76$
TeV. 
The shaded area spanned by two red curves
corresponds to the combination of the mVMD model and the 
Gribov--Glauber model for nuclear shadowing 
with cross section fluctuations
(the shaded area shows the uncertainty of the calculations related to the variation of the fluctuation strength
due to the change of $\beta$ in the range $0.25\leq \beta \leq 0.35$); the blue dashed curve is 
the result of the calculation in mVMD-GM, i.e. without cross section fluctuations;
the green dot-dashed curve is the result of the VMD-DL94 model combined with the Glauber model.
The shape of the rapidity distribution predicted by the mVMD-GGM calculations
is due to specifics of symmetric UPCs and the interplay 
between the energy dependence 
of the inelastic shadowing correction and the 
photon flux. 

The predicted shape of $d\sigma_{Pb Pb \to \rho Pb Pb}/dy$
is different from the almost flat 
$d\sigma_{Pb Pb \to \rho Pb Pb}/dy$
distribution obtained in the VDM-GM and Starlight
approaches and is also
in stark contrast
with the calculations~\cite{Goncalves:2011vf,Santos:2014vwa} 
in the color dipole model approach
predicting a bell-like shape 
for $d\sigma_{Pb Pb \to \rho Pb Pb}/dy$
with the maximum at $y=0$ and small 
values of $d\sigma_{Pb Pb \to \rho Pb Pb}/dy$
at $y \approx - 4.5$ corresponding to $W_{\gamma p}\approx 5-10$ GeV,
i.e., 
to the energy range of the STAR measurements. 
From 
Fig.~\ref{fig:sigma_final_wf_CF_new} it is seen that the 
experimental photoproduction cross
section is almost constant in 
the energy range spanning the STAR and ALICE energies, 
$\sigma_{\gamma Pb\rightarrow \rho Pb}\approx 2$ mb. 
 In UPCs at $y=0$,
 the  
contributions from both colliding nuclei serving as a target are
equal, while at $\mid y\mid =4.5$ 
the contribution of the low energy
photon dominates. 
The photon fluxes
are calculated in all studies similarly and with good accuracy, 
$N_{\gamma /Pb}(y=0)=108$ and $N_{\gamma /Pb}(y=-4.5)=250$. 
Then one easily obtains 
that
$\sigma_{PbPb\rightarrow PbPb\rho}(\mid y\mid =4.5)\approx 500$ mb $>$
$\sigma_{PbPb\rightarrow PbPb\rho}(y=0)\approx 430$ mb. These estimates
confirm that the two-bumped shape 
of the rapidity distribution
seems to be reasonable.

The good agreement with the ALICE result allows us to predict the value of the
cross section of coherent $\rho$ photoproduction in Pb-Pb UPCs 
at $\sqrt{s_{NN}}=5.02$ TeV  in Run 2 at the LHC:
\begin{equation}
\frac {d\sigma (y=0)} {dy} = 560\pm 25 \ {\rm mb} \,.
\end{equation}

Examining 
the calculations of
elastic photoproduction of $\rho$ mesons on nuclei in
the dipole model framework \cite{Goncalves:2011vf,
Santos:2014vwa}, one notes that some of them describe
the STAR and ALICE data while others do not --- the results strongly
depend on the models used for $\rho$-meson wave function and
the dipole cross section.
The dipole model framework
was successfully used in the analyses of 
many 
processes studied at HERA, such as, e.g.,   
deep inelastic scattering (DIS) and vector meson electroproduction in a wide range of the photon virtualities $Q^2$. 
However, in processes dominated by soft physics such as, e.g., in photoproduction of light vector mesons,
the application of the dipole approach is 
subject to
significant theoretical uncertainties including
the need to model the large-size contribution to the dipole cross section and the vector meson wave function.
Considering the CDM predictions for $\rho$ photoproduction in UPCs one finds that it is difficult to describe
simultaneously the $\gamma p \to \rho p$ and $\gamma A \to \rho A$ cross sections.
Note also that the answer is sensitive to  the assumed  effective quark mass which enters in the photon wave function.
Also, the use of a light quark mass 
(for example, $\sim $ 10 MeV in \cite{Santos:2014vwa}) 
in several dipole models leads to a 
very large transverse size of the photon wave function and, consequently, 
to the $t$-dependence of the Compton elastic scattering
which is stronger than that observed in the data~\cite{Breakstone:1981wm}.

\begin{figure}[th]
\begin{center}
\epsfig{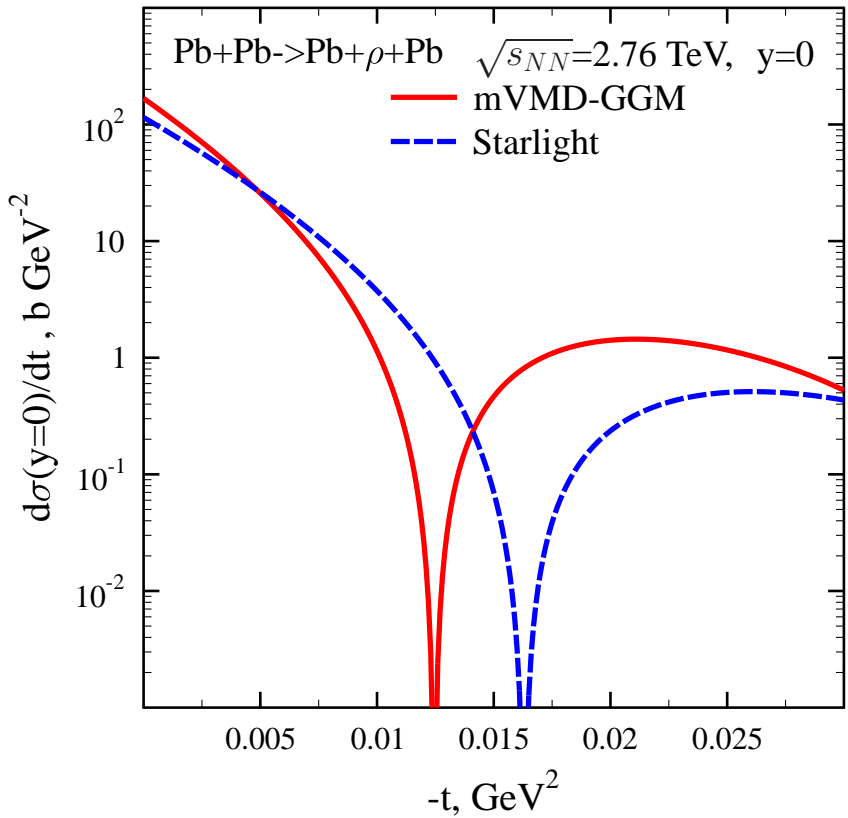}
 \caption{The momentum transfer distribution of coherent $\rho$ photoproduction
in Pb-Pb UPCs at $\sqrt{s_{NN}}=2.76$ TeV. The mVMD-GGM prediction (red solid curve) is compared to the Starlight
result (blue dashed curve).
}
 \label{fig:t_dist}
\end{center}
\end{figure}

We also would like to briefly comment on the description of the 
STAR and ALICE data in the Starlight 
Monte-Carlo generator. 
From our point of view, the observed
agreement with the data on coherent $\rho$ meson photoproduction on nuclei
is 
an effect of lucky coincidence 
and the weak energy dependence
of the photoproduction cross section in the energy range covered
by the STAR and ALICE measurements.
 The Starlight calculations are based on the parametrization of the forward
 $\gamma p\rightarrow \rho p$ cross section, VMD and the optical theorem. 
 The $\gamma A\rightarrow \rho A$ cross
section is calculated in Starlight using the following expression:
\begin{eqnarray}
\sigma_{\gamma A\rightarrow \rho A}=\frac {d\sigma_{\gamma A\rightarrow \rho A}(t=0)}
{dt}\int \limits_{-\infty}^{t_{\rm min}} {\mid F_{A}(t)\mid }^2 dt=
{\frac {1} {16\pi}} \frac {e^2} {f_{\rho}^{2}}
[\sigma_{\rho A}^{\rm tot}]^2
\int \limits_{-\infty}^{t_{\rm min}} {\mid F_{A}(t)\mid }^2 dt \,,
\label{star1}
\end{eqnarray}
where $F_{A}(t)$ is the nuclear form factor normalized by the condition $F_{A}(0)=1$.
Note that the factorized form in Eq.~(\ref{star1}) is an approximation.
While the forward photoproduction cross section on a nuclear target 
in Eq.~(\ref{star1}) follows from the
VMD model, the optical theorem and the Glauber model (see Eq.~(\ref{eq:cs_rho})),
 Eq.~(\ref{star1}) does not take into account that the strong absorption of the $\rho$ meson
in the central region of a heavy nucleus results in narrowing of the momentum
transfer distribution compared to that dictated by the undistorted nuclear
form factor.
  In particular, for heavy nuclei the first 
diffraction minimum is shifted by $10- 15$\% compared to the
position of the dip in the nuclear form factor. This shift is clearly revealed in
the momentum transfer distributions
at the rapidity $y=0$
calculated in mVMD-GGM and 
Starlight approaches, which are shown 
in Fig.~\ref{fig:t_dist}.
The effect of this shift appears to be qualitatively revealed by the STAR (Fig.~2 of \cite{Adler:2002sc})
 and ALICE (Fig.~3 of \cite{Adam:2015gsa}) results.

 A more serious 
 shortcoming of the Starlight calculations is the use of the inelastic $\rho$--nucleus
 cross section instead of the total cross section in Eq.~(\ref{star1}) (compare the
expressions in Eq.~(\ref{haincs})), which violates the optical theorem. 
At high energies the cross section of elastic hadron
scattering by a heavy nucleus is about $30\%$ of the total cross section (see
Fig.~\ref{fig:csha}). As a result, this decreases the estimate of the  forward
cross section by a factor of about two. One can see from 
Fig.~\ref{fig:freecs} 
that the parametrization used in Starlight is close to that given by the DL94 model. 
Therefore,
dividing the value of the VMD-GM cross section (green dot-dashed line) at $y=0$ 
by the factor of two we reproduce the Starlight agreement with the ALICE data.

Finally, we would like to emphasize that our calculations 
use as input the data on the forward $\rho$
photoproduction cross section off the proton and the 
data on  the photon and pion diffraction
on hydrogen in a wide range of energies
$5\,{\rm GeV} \leq W_{\gamma p}\leq 50$ GeV.
As one can see from 
Fig.~\ref{fig:freecs},
the forward $\gamma p\to \rho p$ cross section 
is known 
in the the $5-20$ GeV range 
with rather large experimental errors, while 
 $\sigma_{\rho N}$ extracted
from these data is important for calculation of the cross section 
for the STAR kinematics and for 
predictions
of the  rapidity distribution at large $\mid y\mid $ at the LHC energies. 
Some of these measurements would be doable with the recoil 
detector at COMPASS. Information at higher energies could be 
obtained from the studies of UPCs in $pA$ at the LHC. 
It would be also very important to collect data
on the photon and pion  coherent diffraction off the proton and nuclear targets
at high energies since only a handful of such data 
is
available now. 
In the case of the $\gamma A$ process, one could get this information 
from UPCs at the LHC.

Other possible directions of studies include 
coherent $\phi$ production, where we expect a significant amplification 
of the inelastic intermediate state effects due to the small 
value of $\sigma(\phi N)$ (such a measurement is certainly challenging 
for the main decay channel of $\phi$ into two kaons, 
but the 15\% $\rho \pi$ channel may work). 

The phenomenon of fluctuations of the interaction strength,
which we discussed in the context of $\rho$ 
meson exclusive production on nuclei, should also be manifested in 
a wide range of high energy $\gamma A$ inelastic processes that 
could be studied in UPCs at the LHC.
Effects of such fluctuations in inelastic  $pA$ collisions were
considered in \cite{Alvioli:2013vk} with experimental evidence  reported in
\cite{TheATLAScollaboration:2013cja}.

\section{Conclusion}

With an increase of the collision energy, the composite structure 
of the photon becomes progressively more
pronounced, which leads to the following two features of the 
calculation of the cross section of $\rho$ photoproduction on nuclei 
compared to the lower energies. 
First, the significant cross section of photon inelastic diffraction 
results in the sizable inelastic nuclear shadowing
correction to the $\gamma A \to \rho A$ cross section. 
Second, the QCD-motivated enhancement of 
the hadronic fluctuations of the photon
 reduces the cross section of $\rho$ photoproduction on the proton 
 in agreement with the latest 2006 H1 data. 
We took these features into account by combining the 
modified VMD model 
with the Gribov--Glauber model for nuclear shadowing, 
where the inelastic nuclear shadowing is included by means of 
cross section fluctuations
described using a QCD-motivated parametrization.
The resulting approach allows us to successfully describe 
the data on elastic 
$\rho$ photoproduction on nuclei in heavy ion UPCs 
in the $7 < W_{\gamma p} < 46$ GeV 
energy range and to predict the value of the
cross section of coherent $\rho$ photoproduction in Pb-Pb UPCs 
at 
$\sqrt{s_{NN}}=5.02$ TeV  in Run 2 at the LHC:
$d\sigma_{PbPb \to \rho Pb Pb}(y=0)/dy = 560 \pm 25$ mb.

 \section*{Acknowledgements}

The authors would like to thank S.~Klein and J.~Nystrand for useful comments.


\end{document}